\documentclass[conference]{IEEEtran}

\IEEEoverridecommandlockouts

\usepackage{multirow}

\usepackage{setspace}

\usepackage{amsmath}
\usepackage{amssymb}

\usepackage{algorithmic}

\usepackage{array}

\usepackage{graphicx}

\usepackage[caption=false,font=normalsize,labelfont=sf,textfont=sf]{subfig}

\usepackage{fixltx2e}

\usepackage{stfloats}

\usepackage{url}

\usepackage{lipsum}
\usepackage{marginnote} 
\usepackage{graphicx}
\usepackage[dvipsnames]{xcolor}
\usepackage{tikz}
\usetikzlibrary{backgrounds}
\usetikzlibrary{arrows,shapes}
\usetikzlibrary{tikzmark}
\usetikzlibrary{calc}

\usepackage{amsmath}
\usepackage{amsthm}
\usepackage{amssymb}
\usepackage{mathtools}
\usepackage{wrapfig}
\usepackage{comment}

\usepackage{hyperref}

\hypersetup{colorlinks = true,
	    linkcolor = black, 
	    urlcolor = blue,
        citecolor = blue}

\begin{document}

\title{Privacy-Preserving Detection Method for Transmission Line Based on Edge Collaboration}

\author{\IEEEauthorblockN{Quan Shi*, Kaiyuan Deng\thanks{*Quan Shi is the corresponding author.}
\\School of Information and Communication Engineering
\\University of Electronic Science and Technology of China
\\Chengdu, Sichuan 611731\\
Email\: : shiquan@std.uestc.edu.cn d1571502273@outlook.com}}

\maketitle
\begin{abstract}
Unmanned aerial vehicles (UAVs) are commonly used for edge collaborative computing in current transmission line object detection, where computationally intensive tasks generated by user nodes are offloaded to more powerful edge servers for processing. However, performing edge collaborative processing on transmission line image data may result in serious privacy breaches. To address this issue, we propose a secure single-stage detection model called SecYOLOv7 that preserves the privacy of object detecting. Based on secure multi-party computation (MPC), a series of secure computing protocols are designed for the collaborative execution of Secure Feature Contraction, Secure Bounding-Box Prediction and Secure Object Classification by two non-edge servers. Performance evaluation shows that both computational and communication overhead in this framework as well as calculation error significantly outperform existing works.


\end{abstract}

\section{Introduction}

The paper proposes a privacy-preserving detection model based on edge collaboration to enhance the privacy and security of transmission line detection using object detection. The use of unmanned aerial vehicle (UAV) technology and computer vision has significantly reduced the workload of frontline inspectors, mitigated safety hazards, and enabled automatic detection in transmission line inspection. UAVs equipped with communication devices can achieve high-quality real-time image processing.

However, the current UAV technology for transmission line detection is inadequate to support computationally intensive tasks \cite{1}. The limited computing capability of UAVs and network bandwidth results in high latency and increased energy consumption when executing tasks such as deep neural networks. To minimize computational costs, reduce storage space occupation, decrease computational cost and device power consumption, and reduce communication latency between the UAV terminals \cite{3} and edge nodes, researchers have adopted mobile edge computing technology to offload preprocessed image data and detection tasks of UAVs to trusted third parties for storage and analysis \cite{2}, due to the enormous computational cost of object detection tasks.

It is important to note that transmission line detection images contain a significant amount of sensitive information. When UAVs transmit the original data and tasks to edge nodes, they are unaware of the trustworthiness of these nodes or the potential presence of malicious adversaries in the edge environment. This means that there may be a risk of serious privacy breaches during the transmission and edge computing process for images containing sensitive information. For instance, confidential information such as the location and route of the transmission line, positions of shock absorbers and insulators, and statistical data from electricity meters are all at stake. If hackers manage to steal image information during object detection, it could result in severe damage to company interests and social order.

Protecting sensitive information poses a significant challenge for detecting transmission lines. Existing models primarily focus on small object detection \cite{4}\cite{5}, shape features of fittings \cite{6}, and fault identification \cite{7}, with little attention given to data privacy protection\cite{8}. Furthermore, current data privacy protection methods such as anonymization, obfuscation, and homomorphic encryption \cite{9}\cite{10} have serious drawbacks including reduced data availability and identification accuracy, as well as enormous computational cost. These methods are unsuitable for the complex environment of transmission lines and edge collaboration among UAVs. Therefore, designing a secure and privacy-preserving edge collaboration object detection model for transmission lines is an important research topic.

The contributions of this paper can be summarized as

\subsubsection{}
A series of secure multiparty computation protocols (\textbf{Protocol 3-12}) are designed based on Shamir's secret sharing scheme (\textbf{Protocol 1}) and Gennaro's secret product sharing scheme (\textbf{Protocol 2}). Compared to homomorphic encryption methods, secure multiparty computation has advantages in terms of computation and communication complexity, making it more suitable for object detection scenarios in transmission line.
\subsubsection{}
We propose a secure single-object detection model ($\operatorname{SecYOLOv7}$) based on edge collaboration, aiming to protect the privacy of the object detection process. UAV randomly splits the image data and uploads them to edge nodes separately. Two edge servers invoke the modules of Secure Feature Contraction, Secure Bounding-Box Prediction, and Secure Object Classification at the three stages, to obtain the bounding boxes and classification probabilities of the objects in the image. Since neither party can obtain the complete computation results, $\operatorname{SecYOLOv7}$ can ensure the privacy of the target locations and categories in transmission line.

\subsubsection{}
We prove the security and complexity of the secure computation protocols and $\operatorname{SecYOLOv7}$. Experimental performance evaluation shows that the obtained secure bounding boxes almost completely overlap with those in the plaintext environment, and the time cost is only 2.113s, the communication overhead is 95.15 MB. The computation error of $\operatorname{SecYOLOv7}$ can be maintained at around $10^{-4}$, which has prospective significance for the complex environment of the transmission line and the edge collaboration of UAVs.

The rest of the paper is organized as follows: Section \uppercase\expandafter{\romannumeral2} presents the related work; Section \uppercase\expandafter{\romannumeral3} introduces the preliminaries; Section \uppercase\expandafter{\romannumeral4} describes our model in detail; Section \uppercase\expandafter{\romannumeral5} constructs the $\operatorname{SecYOLOv7}$ framework; Section \uppercase\expandafter{\romannumeral6} formally analyses the model's security and complexity, Section \uppercase\expandafter{\romannumeral7} sets the experiments and evaluates the framework's performance; We conclude the paper in Section \uppercase\expandafter{\romannumeral8}.

\section{Related Work}

Currently, researchers have studied object detection models for transmission line using edge computing. Although two-stage object detection algorithms such as Faster R-CNN \cite{11} demonstrate high accuracy, their slow processing speed makes them unsuitable for large-scale tasks. However, some researchers have applied these algorithms in component recognition and defect detection of transmission line components like equalization rings and seismic hammers \cite{12} or nest detection \cite{12}. Additionally, Hu \cite{122} developed a lightweight algorithm specifically designed for UAV edge computing to improve the classification accuracy and real-time performance of the Faster R-CNN algorithm in transmission tower identification.

In contrast, edge computing detection algorithms primarily focus on one-stage object detection for faster performance. Examples include YOLO \cite{13} and SSD \cite{14}, which are trained end-to-end without prior region proposal generation. Ohta \cite{15} used a UAV with YOLO to detect power transmission towers at high locations, while Huang \cite{155} integrated edge computing with YOLOv5s for real-time transmission line detection and improved accuracy of blurry fault target images.

During data processing, UAVs transfer computationally intensive tasks to edge servers, which raises privacy concerns. Currently, researchers are exploring secure multiparty computation schemes based on homomorphic encryption (HE) \cite{16} for processing sensitive data. However, HE significantly increases computational cost\cite{17} in intensive encryption operations and is unsuitable for edge computing scenarios.

Another approach to preserving data privacy while applying data mining techniques is differential privacy (DP) \cite{18}. However, DP-based schemes often compromise model accuracy. Ruan \cite{19} proposed a DPSGD method for secure multiparty computation that strikes a balance between privacy, efficiency, and accuracy to some extent. Huang \cite{20} designed a CNN framework embedded with additive secret sharing for mobile sensing; however, it fails to conceal target locations.

\section{Preliminaries}
\subsection{YOLOv7}

$\operatorname{YOLOv7}$ \cite{21} is an object detection algorithm with three modules: backbone, neck, and head. The backbone module includes convolutional layers, E-ELAN, MP, and SPPCSPC modules to expand the network's learning capacity while balancing computational complexity and accuracy. The neck module fuses multi-level feature maps for improved object detection accuracy. Finally, the head network uses these features for locating and classifying objects with bounding boxes while NMS refines predicted boxes for better performance. It's more efficient than prior versions.

\subsection{Secure Multi-Party Computation}

Secret sharing \cite{23} is an important means of information security and one of the basic application technologies in secure multiparty computation and federated learning. It uses a threshold scheme to divide the secret $\mathcal{S}$ into several shares, which are distributed among participants $\mathcal{P}=\{\mathcal{P}_1,\mathcal{P}_2,...,\mathcal{P}_N\}$. Only ${k}$ or more shares held by participants can reconstruct $\mathcal{S}$, while fewer than ${k}$ shares held by participants cannot reconstruct $\mathcal{S}$, and other subsets cannot obtain any useful information about the secret $\mathcal{S}$.

Currently, three efficient MPC protocols are used based on three Linear Secret Sharing Schemes: Additive Secret Sharing, Shamir's Secret Sharing, and Complementary Nonlinear Function (CNF) Secret Sharing \cite{24}. Additive Secret Sharing is mainly for dishonest majority conditions while Shamir's and CNF Secret Sharing are for honest majority conditions. In the edge computing scenario of transmission line detection, we adopt Shamir's method due to its high computational efficiency.

\begin{figure}[htbp]
    \centering
    \includegraphics[width=\linewidth]{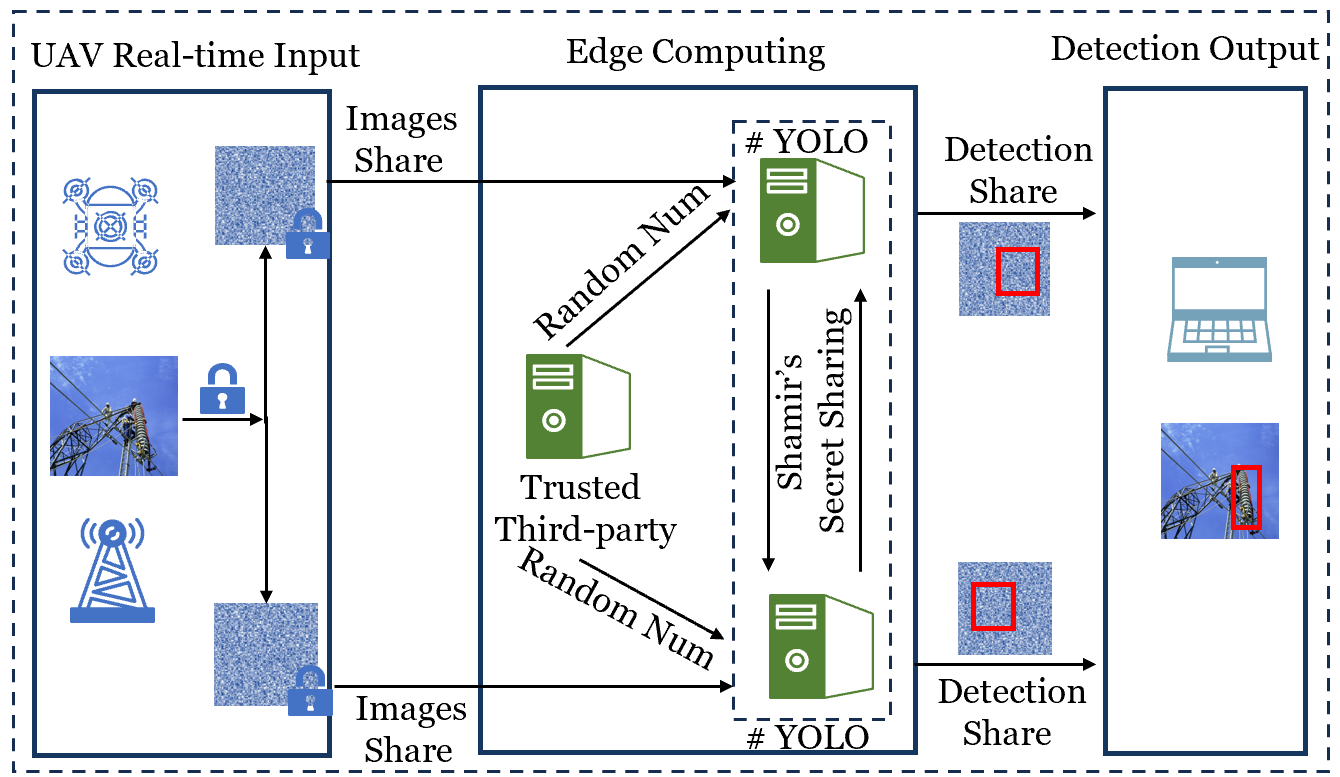}
    \caption{System Model Of SecYOLOv7 framework}
    \label{1}
\end{figure} 

\subsection{Shamir's Secret Sharing}

For Shamir's Secret Sharing \cite{23}, a secret $x$ is first used to generate a polynomial $f(y) = a_0 + a_1y_1+ a_2y_2 +...+ a_t y_t$, where $a_0=x$, and $a_1...a_t$ are random coefficients. The value of $t+1$ represents the minimum number of parties required to reconstruct the secret $y$. Subsequently, we construct $m$ points using indices ranging from $1$ to $m$ from this polynomial. These points, denoted as shares of the secret $x$, are precisely represented by $f(i)$. To reconstruct $x$, all or some subset of parties exchange $f(i)$ and use Lagrange interpolation to recover the polynomial $f(y)$, thereby obtaining $x$. In this paper, we refer to it as \textbf{Protocol 1: SSAdd}, which enables secure computation of $\sum{[x]}$.

Numerous secure distributed multiplication protocols based on diverse subprotocols have been continuously developed for the purpose of MPC, owing to its significance. Secret sharing facilitates collaborative calculation of the confidential product without disclosing any additional information about the secret, as participants can utilize their respective information fragments while computing the product involving the secret input. This procedure entails each party locally multiplying their input shares $[x]_i$ and disseminating the outcomes to all other parties (thus reducing the polynomial degree from $2t$ to $t$). Subsequently, locally computing the linear combination of these shares serves as their output share $[y]$.

To improve the multiplication's efficiency and feasibility, Gennaro \cite{25} proposed a two-party secret multiplication sharing scheme based on the Shamir's secret sharing scheme \cite{23}. Given polynomials $h_1(x), \ldots, h_{2 t+1}(x)$ all of degree $t$ which satisfy that $h_i(0)=f_{\alpha \beta}(i)$ for $1 \leq i \leq 2 t+1$, and define $H(x) \stackrel{\text { def }}{=}$ $\sum_{i=1}^{2 t+1} \lambda_i h_i(x)$. Note that $H(0)$ is $\lambda_1 f_{\alpha \beta}(1)+\ldots+$ $\lambda_{2 t+1} f_{\alpha \beta}(2 t+1)$ and hence $\alpha \beta$. What's more, $H(j)=\sum_{i=1}^{2 t+1} \lambda_i h_i(j)$. This method greatly improves the efficiency. In this paper, we use Gennaro's multiplication sharing scheme as \textbf{Protocol 2: GRRMult}, to safely computes $[z]=\prod{[x]}$.

\section{System Description}

\subsection{System Model}

In the proposed framework, there are several main components: UAV sender $I$, which captures images of the transmission line and randomly splits them into two parts $I_1$ and $I_2$; two edge servers $\mathcal{P}_1$ and $\mathcal{P}_2$; a trusted third-party server $\mathcal{T}$, as shown in Figure 2; and image receiver $O$.

Trusted third-party $\mathcal{T}$: Serving as a trusted secret distribution entity, $\mathcal{T}$ is responsible for generating encryption keys used for image encryption and securely transferring them to the UAV sender $I$.

Input $I$: The UAV acts as the image capture device equipped with a high-resolution camera. It captures images of critical sections of the transmission line and generates image data. Each image $I$ is linearly encrypted using $\operatorname{SSAdd}$, creating image subsets $I_i$, which are separately sent to $\mathcal{P}_i$. This process ensures the privacy protection of the image data.

Participant $P_i$: The edge servers $P_i$ employ the $\operatorname{SecYOLO v7 }$ model, which follows a series of security protocols to obtain two classification results $O_i$ and send them to $O$.

Output $O$: Upon receiving $O_i$, $O$ performs reconstruction decryption (Shamir's linear reconstruction decryption) \cite{23} to efficiently recover the original classification results.
\subsection{Security Model}

This paper investigates techniques for preserving the privacy of encrypted images during the inference process. The proposed model ensures robust security measures, limiting access to the original image, extracted features, and final classification results solely to authorized parties - sender and recipient. As a result, external entities are prevented from accessing valuable image information.

We adopt an Honest-But-Curious (HBC) security model, assuming the honesty, curiosity and independence of all involved parties including $\mathcal{P}_i$ and the trusted third party $\mathcal{T}$. Although attackers possess both intention and capability to harm participating entities, they will adhere to protocols without coordination or causing damage. The trusted third party $\mathcal{T}$ is solely responsible for offline generation or distribution of encryption keys and refrains from computations or accessing any grid image data.

Within the HBC model, numerous potential adversaries endeavor to extract image privacy information, which is similar to \cite{19}.

\section{Construction Of SecYOLOv7 Framework}

\subsection{Model Overview}

\begin{figure*}[htbp]
    \centering
    \includegraphics[width=\linewidth]{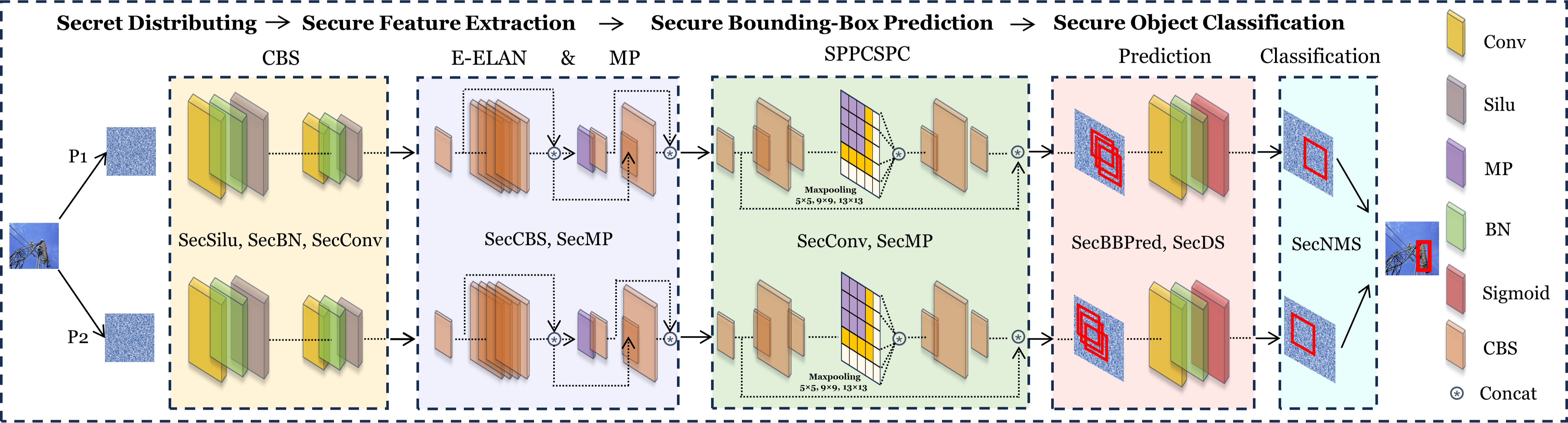}
    \caption{Overview Of SecYOLOv7 Network}
    \label{2}
\end{figure*} 
Figure 2 illustrates the framework of $\operatorname{SecYOLOv7}$. The original image is randomly partitioned into two segments and uploaded to the $\mathcal{P}_i$ device for collaborative execution of Secure Feature Contraction, Secure Bounding-Box Prediction, and Secure Object Classification protocols. The Secure Feature Contraction module employs collaborative execution steps such as CBS, E-ELAN, MP, and SPPCSPC to augment receptive field and feature representation using protocols like $\operatorname{SecSilu}$, $\operatorname{SecBN}$, $\operatorname{SecConv}$, and $\operatorname{SecMP}$. In the Secure Bounding-Box Prediction module, we use integrated protocols such as $\operatorname{SecBBPred}$ with $\operatorname{SecSd}$ to predict bounding boxes and their associated probabilities. The Secure Object Classification module involves candidate box filtering using embedded $\operatorname{SecDS}$ in conjunction with $\operatorname{SecNMS}$. Finally, genuine object boundaries and classification outcomes can be reconstructed from information contained in both partitions.

\subsection{Basic Computing Protocols}

Based on Shamir's secret sharing scheme \cite{23} (\textbf{Protocol 1}) and Gennaro's secret product sharing scheme (\textbf{Protocol 2}), three basic secure computation protocols, namely $\operatorname{SecDivi, SecComp,}$ and $\operatorname{SecExp}$, have been designed. The details are provided in  \href{https://github.com/Ogatarina-sq/secyolov7/tree/main/missingprotocols}{https://github.com/Ogatarina-sq/secyolov7/tree/main/missingprotocols}\textbf{, Protocol 3-5}. The protocols can be summarized as follows:

\textbf{Protocol 3: } Secure Division (SecDivi) Protocol:

$\mathcal{P}_1$ has shares $[u]^{\prime}$, $[v]^{\prime}$ ; $\mathcal{P}_2$ has shares $[u]^{\prime \prime}$, $[v]^{\prime \prime}$ ;  $\mathcal{P}_i$ jointly computes $u=[u]^{\prime}+[u]^{\prime \prime}$, $v=[v]^{\prime}+[v]^{\prime \prime}$ .$[y]^{\prime}$ , $[y]^{\prime \prime}$ are two random shares of the division result, i.e. $[y]^{\prime}+[y]^{\prime \prime}=\frac{v}{u}$.

\textbf{Protocol 4: } Secure Comparison (SecComp) Protocol:

$\mathcal{P}_1$ has shares $[u]^{\prime}$,$[v]^{\prime}$,$\mathcal{P}_2$ has shares $[u]^{\prime \prime}$,$[v]^{\prime \prime}$;$\mathcal{P}_1$ and $\mathcal{P}_2$ jointly computes the sign $[y]$,if $[u]^{\prime}+[v]^{\prime}\geq[u]^{\prime \prime}+[v]^{\prime \prime}$,$[y]=0$;i.e. if $[u]^{\prime}+[v]^{\prime}\textless[u]^{\prime \prime}+[v]^{\prime \prime}$,$[y]=1$.

\textbf{Protocol 5: } Secure Exponentiation (SecExp) Protocol:

$\mathcal{P}_1$ has shares $[u]^{\prime}$ ; $\mathcal{P}_2$ has shares $[u]^{\prime \prime}$ ; $\mathcal{P}_i$ returns $[y]^{\prime}$ , $[y]^{\prime \prime}$ ,i.e. $[y]^{\prime}+[y]^{\prime \prime}=e^{[u]^{\prime}+[u]^{\prime \prime}}$.

\subsection{Secure Feature Extraction}

Image feature extraction is a necessary step in Object Detection and serves as the Backbone layer in the YOLO framework. In the YOLO V7 algorithm, the modules used for image feature extraction are the CBS (Convolutional Block Structures) layer, E-ELAN layer, and MP (Max Pooling)layer. 

\subsubsection{Secure CBS}

After the image is input into $\operatorname{YOLOV7}$, the first step involves feature extraction and downsampling, which begins with the CBS module. The network consists of three different modes of the CBS module, each utilizing distinct convolution kernels $(k)$ and strides $(s)$. CBS1 employs a $1\times1$ convolution with a stride of 1 to modify channel numbers. CBS2,3 are $3\times3$ with 1,2 strides. But all of them pass through Conv, BN, and Silu modules

To address this, we have designed the Secure Convolution Module $\operatorname{(SecConv)}$, Secure Normalization Module $\operatorname{(SecBN)}$, and Secure Silu Activation Function $\operatorname{(SecSilu)}$.

The Conv layer is linear and involves matrix dot product, making it compatible with $\operatorname{SSAdd}$. 2D convolution can be defined as $Conv(x,(w,b))=\bar{x} w+b$, where $x$ represents the feature matrix, $w$ represents the kernel weights, and $b$ represents the bias. When the kernel slides over $x$, resulting in a convolution result of $[s]$, sub-matrices in $\bar{x}$ are dotted with $w$.

Our designed $\operatorname{SecConv}$ protocol achieves secure convolution operations by employing $\operatorname{SSAdd}$ and $\operatorname{GRRMult}$. Specifically, $x$ is randomly split into two shares, $[x]_i$, and $\operatorname{GRRMult}$ protocol is adopted to securely compute $[s]_i = \operatorname{SSAdd}(\operatorname{GRRMult}([x]_i, (w)), b)$ for each share. Finally, the final convolution result $[s]$ is obtained as $[s]=[s]_1+[s]_2$.

\begin{math}
\begin{aligned}
\hline&\textbf{Protocol 6:}\text{ SecConv: Secure Conv Protocol}\\
\hline&\textbf{Input: } \mathcal{P}_i \text{ has shares }[x]_i,  \text{ kernel weight }{w},  \text{ bias }{b}\\
&\textbf{Output:} \mathcal{P}_i \text{ returns shares }[s]_i\\
&1) \mathcal{P}_i \text{ jointly }[s]_i=\operatorname{SSAdd}\left(\operatorname{GRRMult}\left(\overline{[x]_i},(w)\right)\right., b)\\
&2) \mathcal{P}_i \text{ returns } [s]_i, \text{ for } y=[s]_1+[s]_2=\operatorname{Conv}(x,(w, b))\\
\hline\\
\end{aligned}
\end{math}

We have developed the Secure Batch Normalization $\operatorname{(SecBN)}$ protocol to implement the functionality layer of BN, as described in the $\operatorname{SecBN}$ protocol. The $\operatorname{SecBN}$ protocol enables party $P_i$ to obtain the normalized and linearly transformed feature share $[y]_i$ from their respective feature share $[x]_i$, for each input feature $[x]_i$ and various parameters involved in batch computations in deep learning algorithms (mean $\mu$, variance $\sigma^2$, scale $\gamma$, and shift $\beta$). This process facilitates the reconstruction of the feature vector $y$. Mathematically, it can be represented as $y=\gamma \cdot x+\beta=\gamma \cdot \frac{x-\mu}{\sqrt{\sigma^2+\varepsilon}}+\beta$. The resulting feature share $[y]_i$ produced by the $\operatorname{SecBN}$ protocol can then serve as input for subsequent $\operatorname{SecSilu}$ operations.

\begin{math}
\begin{aligned}
 \hline &\textbf { Protocol 7:}  \text{ SecBN: Secure Batch Normalization Protocol} \\
 \hline &\textbf { Input: } \mathcal{P}_i \text { has shares }[x]_i \text { and }
 \text { parameters : mean } \mu,\\ &\quad \text { variance } \sigma^2, \text { scale } \gamma \text {, shift } \beta . \\
& \textbf { Output: } \mathcal{P}_i \text { output shares }[y]_i \text { after secure BN  } \\
& 1) \mathcal{P}_i \text { locally computes normalized }[\overline{x}]_i \leftarrow \frac{[x]_i-\mu / 2}{\sqrt{\sigma^2+\varepsilon}} ; \\
& 2) \mathcal{P}_i \text { locally computes transformed  } 
 {[y]_i \leftarrow \gamma \cdot[\overline{x}]_i+\beta / 2 ;} \\
& 3) \mathcal{P}_i \text { returns }[y]_i; \\
\hline\\
\end{aligned}
\end{math}

The SiLU activation function, used in the CBS module, is a smoother alternative to the traditional ReLU. It maps input values to the range of $[0, 1]$ and enhances network fitting capability for better image feature capture and improved object detection accuracy. Our designed \textbf{Protocol 8: } $\operatorname{SecSilu}$ protocol achieves secure exponentiation, addition, and division operations using $\operatorname{SecExp, SSAdd, }$ and $\operatorname{SecDivi}$ functions.\\

\begin{math}
\begin{aligned}
\hline &\textbf{Protocol 8: } \text{ SecSilu: Secure Silu Protocol } \\
\hline &\textbf{Input: } \mathcal{P}_i \text{ has shares }[x]_i \\
&\textbf{Output: } \mathcal{P}_i \text{ outputs shares }[y]_i\\
&1)  \mathcal{P}_i \text{ jointly }[e]_i\leftarrow\operatorname{SecExp}([x]_1,[x]_2)\\
&2)  \mathcal{P}_i \text{ jointly }[s]_i\leftarrow\operatorname{SSAdd}([x]_1,[x]_2)\\
&3)  \mathcal{P}_i \text{ jointly }[y]_i\leftarrow\operatorname{SecDivi}([s]_1,[s]_2,1-[e]_1,-[e]_2)\\
&4)  \mathcal{P}_i \text{ returns }[y]_i\\
\hline\\
\end{aligned}
\end{math}

If we turn $[s]_1,[s]_2$ from SecSilu's 3) into $1$ and $0$, it becomes Secure Sigmoid Protocol (SecSd).

To sum up, we are able to construct the Secure CBS (SecCBS) model with $\textbf{Protocol 5-7}$ embedded.

\subsubsection{Secure MP}

The MP module consists of two branches for downsampling. The main objective of the \textbf{Protocol 9: } $\operatorname{SecMP}$ layer is to calculate the maximum value within a $2\times2$ grid: $max[x(j, k), j, k = 0,1]$. Its purpose is to prevent sensitive information disclosure to $\mathcal{P}_i$, while optimizing array utilization as computational units to minimize additional overhead associated with multi-level loops. For each pooling sub-region of size ($n, n$), $\mathcal{P}_i$ reshapes the corresponding feature shares $[x]_i$ into a $(1,n^2)$ matrix and then applies $\operatorname{SecComp}$ and $\operatorname{GRRMult}$ protocols for $n^2-1$ rounds to obtain the maximum feature share $[y]_i$.

\begin{math}
\begin{aligned}
\hline&\textbf{Protocol 9:} \text{SecMP: Secure Max-Pooling Protocol}\\
\hline&\textbf{Input: }\mathcal{P}_{i}\text{ has }[\vec{x}]_{i}; \\
&\textbf{Output: }\mathcal{P}_{i} \text{ returns } [y]_{i};\\
&1) \mathcal{P}_{i} \text{ inits } [y]_{i}\leftarrow[x_0]_{i}; \\
&2)\textbf{  for } j=1 \text{ to } n^2-1 \textbf{ do } \\
&\left\{
\begin{aligned}
& \mathcal{P}_{1} \text{ and } \mathcal{P}_{2} \text{ compute } 
[f]_{1},[f]_{2}\leftarrow \\& \operatorname{SecComp}([y]_{1},[y]_{2},[x_{j}]_{1},[x_{j}]_{2}); 
[y^{\prime}]_{1},[y^{\prime}]_{2} \leftarrow \\&  \operatorname{GRRMult}([x_{j}]_{1}-[y]_{1},[x_{j}]_{2}-[y]_{2},[f]_{1},[f]_{2}); 
\end{aligned}
\right. \\
&3) \mathcal{P}_i \text{ locally computes }[y]_i \leftarrow[y]_i+[y^{\prime}]_i;\\
&4) \mathcal{P}_i \text{ returns }[y]_i.\\
\hline\\
\end{aligned}
\end{math}

\subsubsection{Secure E-ELAN}

The E-ELAN architecture in $\operatorname{YOLOv7}$ uses group convolution, shuffle operation, and merge cardinality to improve learning capability of feature maps, optimize parameter utilization, enhance computational efficiency and network robustness without altering gradient propagation path.

We have devised the Secure E-ELAN protocol by incorporating the $\operatorname{SecCBS}$ protocol. The Secure E-ELAN module is formed by concatenating $\operatorname{SecCBS1}(Line1)$, $\operatorname{SecCBS1}$, 
[$\operatorname{SecCBS2}$]$\times$2$(Line2)$, and $\operatorname{SecCBS1,[SecCBS2]}\times$4$(Line3)$ along the channel dimension to generate a novel feature map. Its correctness and security can be easily demonstrated.

\subsubsection{Secure SPPCSPC}

One improvement of $\operatorname{YOLOv7}$ over YOLOv5 is the addition of the SPPCSPC module, an enhanced version based on the SPP module. It extracts feature maps of different scales $(5\times5, 9\times9, 13\times13)$ from intermediate layers using max pooling operations and combines them to generate a multi-scale feature vector. This enhances object detection accuracy by capturing object features at various scales while reducing computational complexity. The Secure SPPCSPC module consists of {$\operatorname{SecConv, SecMaxPool, SecConv}(line\:k)$} $(k=1,2,3)$, which are concatenated along the channel dimension to create a new feature map.

In summary, we are able to construct the Secure Feature Extraction model with Secure CBS, Secure MP, Secure E-ELAN and Secure SPPCSPC embedded.

\subsection{Secure Bounding-box Prediction}

\begin{math}
\begin{aligned}
\hline&\textbf{Protocol 10:} \text{ Secure Bounding-Box Prediction (SecBBPred)}\\
\hline&\textbf{Input:} \mathcal{P}_i \text{ has shares }[x]_i, \text{ K-means cluster centers } C, \\&\text{ kernel weights } w, \text{ bias }b, \text{ BN parameters }\mu, \sigma^2, \gamma, \beta\\
&\textbf{Output:} \mathcal{P}_i \text{ outputs shares }[y]_i\\
&1) \textbf {for } j=1 \text { to end }\textbf{do } 2)- 9)  \\
&2) \mathcal{T} \text { generates random } \bar{u}_{1, j} ,\bar{u}_{2, j} \in \mathbb{Z}, \bar{u}_{i, j} \leftarrow \bar{u}_{1, j}+\bar{u}_{2, j}; \\
&3) \mathcal{P}_1 \text { computes } \varphi_j \leftarrow [x]_{1, j}-\bar{u}_{1, j}, \mathcal{P}_2 \text { sends }\lambda_j \leftarrow \\ &[x]_{2, j}-\bar{u}_{2, j}\text { to } \mathcal{P}_1.  \mathcal{P}_1 \text { sends } \bar{P}_j \leftarrow \varphi_j-\lambda_j \text { to }  \mathcal{P}_2; \\
&4) \text { if } \bar{P}_j<0 , \mathcal{P}_i \text { computes } U_{i, j} \leftarrow [x]_{i, j};\text{ else } U_{i, j} \leftarrow \bar{u}_{i, j}\\
&5)\mathcal{P}_i \text{ locally computes } U_{i, j}^{\prime} \leftarrow [x]_{i, j}+u_{i, j}-U_{i, j};\\
&6)\mathcal{P}_i \text{ jointly } O_{i, j} \leftarrow \operatorname{SecDivi}\left(U_{1, j}, U_{2, j}, U_{1, j}^{\prime}, U_{2, j}^{\prime}\right);\\
&7)\mathcal{P}_i \text{ locally computes } d_{i, j} \leftarrow 0.5-O_{i, j};\\
&8)\mathcal{P}_i \text{ initializes } dc_{i, j} \in \mathbb{Z} ,\mathcal{P}_i \text{ computes } dv_{i, j} \leftarrow d_{i, j};\\
&9) \textbf{ for } cs=1 \text{ to }\varpi\:\mathcal{P}_i  \textbf{ do } \\
&\left\{
    \begin{aligned}
& V_{1, j, c P}, V_{2, j, c P} \leftarrow \operatorname{SecComp}\left(dv_{i, j, c P},d_{i, j, c P}\right); \\
&V_{i, j, c P}^{\prime} \leftarrow \frac{V_{i, j, c P}+2-i}{2};\\
& V_{i, j, c P}^{\prime \prime} \leftarrow \operatorname{GRRMult}\left(d_{i, j, c P}-d v_{i, j, c P},  V_{i, j, c P}^{\prime}\right); \\
& [\text{cluster}]_{i, j, cP} \leftarrow V_{i, j, cP}^{\prime \prime}+1+d v_{i, j, cP}
\end{aligned}
\right. \\
&10) \quad I_i = \text{ExtractIndices}([x]_i, [\text{cluster}]_i) \\
&11)  \quad [c_1], [c_2], ..., [c_k] = \text{GetClusterCenters}(C) \\
&12)  \quad [d_j]_i = [x]_i - [c_j], \quad \forall j \in I_i \\
&13) \quad [\text{center}]_i = \frac{\text{SSAdd}_{j \in I_i} [d_j]_i}{|I_i|} \\
&14) \mathcal{P}_i \text{ computes }[s]_i = \text{SecConv}([\text{center}]_i, {w}, {b}) \\
& [z]_i = \text{SecBN}([s]_i, \mu, \sigma^2, \gamma, \beta),[y]_i = \text{SecSd}([z]_1, [z]_2) \\
&15)\mathcal{P}_i \text{ returns } [y]_i \\
\hline\\
\end{aligned}
\end{math}

The $\operatorname{SecYOLOv7}$ model uses secure and adaptive $\operatorname{K-means}$ clustering to generate bounding boxes by grouping similar sample boxes together based on their distances, computing cluster centers from the average values of samples within each cluster, and repeating this process iteratively until obtaining the final set of bounding boxes.

The Secure Bounding-Box Prediction process, as described in \textbf{Protocol 10}, involves the following steps. Let $[x]_i$ represent the input sample boxes, and let $w$ denote the number of clusters. Initially, $\mathcal{P}_i$ sets the number of clusters and randomly initializes the cluster centers, $u_i$. Then, $\mathcal{P}_i$ calculates the distances between $u_i$ and the input boxes $[x]_i$, using a distance metric $d=1-IoU$, where $IoU$ represents the Intersection over Union \cite{155}, which measures overlap between two anchor boundaries. After that, $\operatorname{SecBBPred}$ uses a secure Sigmoid activation $\operatorname{SecSd}$ to perform secure prediction tasks for multiple objects and labels. After this step, we need to handle those forecasted boxes.
\subsection{Secure Object Classification}

\begin{math}
\begin{aligned}
\hline&\textbf{ Protocol 11:}\text{  SecDS : Secure Descending Sort Protocol }\\
\hline&\textbf{ Input: } \mathcal{P}_1 \text{ has } [u]_1 \in \mathbb{Z}_n, \mathcal{P}_2 \text{ has } [u]_2 \in \mathbb{Z}_n\\
&\textbf{ Output: }\mathcal{P}_1 , \mathcal{P}_2 \text{ outputs } s\\
&1) \mathcal{P}_i \text{ generates random } \rho_i \in \mathbb{Z}_n;\\
&2) \mathcal{P}_i \text{ computes }[v]_i \leftarrow [u]_i-\rho_i , \text{ sends }[v]_i \text{ to }\mathcal{P}_{3-i};\\
&3) \mathcal{P}_i \text{ jointly computes }[v] \leftarrow \operatorname{SSAdd}([v]_1,[v]_2)\\
&4) \mathcal{P}_i \text{ descending }\operatorname{sort}([v])\\
&5) \mathcal{P}_i \text{ returns }s \\
\hline\\
\end{aligned}
\end{math}

\begin{math}
\begin{aligned}
\hline&\textbf{Protocol 12:} \text{ SecNMS Protocol }\\
\hline&\textbf{Input: } \mathcal{P}_1 \text{ has } U_1 \in \mathbb{Z}_n , \mathcal{P}_2 \text{ has } U_2 \in \mathbb{Z}_n ,\text{  threshold }\eta^{\prime}\\
&\textbf{Output: } \mathcal{P}_1 \text{ returns }V_1, \mathcal{P}_2 \text{ returns } V_2\\
&1) \mathcal{P}_i \text{ inits } U_i \leftarrow\left\{U_i^{x^{\prime}}, U_i^{y^{\prime}}, U_i^{x^{\prime \prime}}, U_i^{y^{\prime \prime}}\right\}\\
&2) \mathcal{P}_1 \text{ sends } U_1^w \leftarrow U_1^{x^{\prime \prime}}-U_1^{x^{\prime}}\text{ \& }U_1^h \leftarrow U_i^{y^{\prime \prime}}-U_1^{y^{\prime}}\text{ to }\mathcal{P}_2\\
&3) \mathcal{P}_2 \text{ computes }U_2^w \leftarrow U_2^{x^{\prime \prime}}-U_2^{x^{\prime}} \text{ and } U_2^h \leftarrow U_i^{y^{\prime \prime}}-U_2^{y^{\prime}},\\& \quad\text{ sends }U_2^w \text{ and }U_2^h \text{ to }\mathcal{P}_1\\
&4) \mathcal{P}_i \text{ computes }S \leftarrow\left(U_1^w+U_2^w\right)\left(U_1^h+U_2^h\right)\\
&5) \mathcal{P}_i \text{ jointly computes }\gamma \leftarrow \operatorname{SecDS}\left(P_1, P_2\right)\\
&6) \mathcal{P}_i \text{ creating index lists } \Psi \\
&7) \textbf{ where }\gamma \neq \varnothing,\\
&\left\{
    \begin{aligned}
&\mathcal{P}_i \text{ inits } \partial\leftarrow Y[0],\text{ add }\sigma \text{ to the end of } \\& \Psi , \text{ delete }\sigma \text{ in }\gamma \\
 &\textbf{ for }k \in \gamma,\textbf{ do }\\
&\left\{
    \begin{aligned}
&\mathcal{P}_i \text{ computes }T_i \leftarrow U_i[(i-1)\partial] ;\\
&p^{\prime} \leftarrow \operatorname{SecComp}\left(U_1^{x^{\prime}}[\partial]\right.,\left.U_2^{x^{\prime}}[\partial], U_1^{x^{\prime}}[k], U_2^{x^{\prime}}[k]\right)\\
 &\text{if } p^{\prime}=1,
\mathcal{P}_i \text{ jointly }T_i^{x^{\prime}} \leftarrow U_i^{x^{\prime}}[k],\left\{T_i^{y^{\prime}}, T_i^{x^{\prime \prime}}, T_i^{y^{\prime \prime}}\right\}\\
&\mathcal{P}_i \text { jointly computes }T_1+T_2 \text{ area } s\\
 & \text{ and }s^{\prime} \leftarrow max(s, 0) \text{ and }\mathrm{IoU} \leftarrow \frac{s^{\prime}}{S[\sigma]+S[k]-s^{\prime}}\\
& \text{ if } \mathrm{IoU} \geqslant \eta^{\prime}, \mathcal{P}_1 \text{ and }\mathcal{P}_2 \text{ delete }k \text{ in} \Upsilon \\
\end{aligned}
\right. \\
\end{aligned}
\right. \\
&13) \mathcal{P}_i \text{ returns }V_i \leftarrow U_i(\Psi).\\
\hline\\
\end{aligned}
\end{math}

Prior to conducting Non-Maximum Suppression (NMS), it is imperative to establish a Secure Descending Sorting ($\operatorname{SecDS}$) protocol, as delineated in \textbf{Protocol 11}. Given replicas of bounding box probabilities $[u]_i$, sorting algorithms like quicksort can be employed by $\mathcal{P}_i$ to arrange $[v]$ in descending order, thereby obtaining a descending index list for $[u]_i$.

To enhance detection efficiency while maintaining accuracy, $\operatorname{SecYOLOv7}$ employs the non-maximum suppression (NMS) method. We propose the $\operatorname{SecNMS}$ protocol to securely select a set of bounding boxes with high detection scores and low redundancy. Given duplicate target bounding boxes $U_i$ and their corresponding probability duplicates $\mathcal{P}_i$, we compute the area $S$ of the bounding boxes using coordinate differences without revealing complete coordinate values. Then, $\mathcal{P}_i$ invokes the $\operatorname{SecDS}$ protocol to calculate an index list $\gamma$ in descending order of probability values, retaining only the bounding box index with highest probability. The remaining bounding boxes are compared for similarity to this retained one by computing overlapping areas through secure computation. Bounding box similarity is measured by $(IoU)$, which is defined as ratio of intersection area $A\cap B$ to union area $A\cup B$. If $\frac{S(A\cap B)}{S(A)+S(B)- S(A\cup B)}\geq\eta^{\prime}$, $A$ and $B$ are considered similar where $\eta$ is similarity suppression threshold. For indices $\sigma$ and $k$, overlap region $s$ is computed; if s$\geq0$, two bounding boxes are considered non-overlapping resulting in $IoU=0$; else redundant indices are removed from list $\gamma$. This process iteratively classifies bounding box indices until termination when  $\gamma<\varnothing$. Based on index list obtained,  duplicate bounding box $V_i$ suppressed by $\operatorname{SecNMS}$ can be determined.\\

\section{Security Analysis And Complexity}

\subsection{Security Analysis}

The security analysis of the proposed protocol is presented in this section, aiming to demonstrate the feasibility of simulating the corrupted party's perspective based on its input and output. To achieve this, $\operatorname{Lemma1-3}$ is referenced.

\textbf{Definition 1:} The Honest-But-Curious (HBC) model defines a secure protocol as one for which there exists a probabilistic polynomial-time (PPT) simulator $S$ capable of generating simulation views for the adversary $A$ in the real world, such that these views are computationally indistinguishable from their genuine counterparts.

\textbf{Lemma 1:} \cite{26} The protocol is considered to be perfectly simulatable if all of its sub-protocols are also deemed perfectly simulatable.

\textbf{Lemma 2:} \cite{27} If $[u]_1$ and $[u]_2$ are indistinguishable for uniformly random elements $[u]_1, [u]_2\in \mathbb{Z}_n$, then linear operations on $[u]_1$ and $[u]_2$ also yield uniformly random results that cannot be distinguished from either $[u]_1$ or $[u]_2$.

\textbf{Lemma 3:} \cite{23}\cite{28} The $\operatorname{SSAdd}$ protocol and the $\operatorname{GRRMult}$ protocol are secure within the HBC model.

\textbf{Theorem 1:} The $\operatorname{SecDivi, SecComp}$, and $\operatorname{SecExp}$ protocols are secure within the HBC model.

The security of the sub-protocols $\operatorname{SecDivi}$, $\operatorname{SecComp}$, $\operatorname{SecExp}$ has been formally proven in \cite{19}. These protocols operate on arrays as inputs and due to the uniform and random distribution of these inputs, it can be shown that the resulting arrays generated during computation are also uniformly distributed based on $\operatorname{Lemma 2}$.

\textbf{Theorem 2:} The $\operatorname{SecConv, SecSilu, SecSd, SecBN, SecMP}$, $\operatorname{SecBBPred, SecDS,}$ and $\operatorname{SecNMS}$ protocols are secure within the HBC model.

Taking $\operatorname{SecSilu}$ as an example, the real view of $ \mathcal{P}_1 $ consists of $([x]_1, [e]_i, [s]_i, [y]_1)$, while the real view of $\mathcal{P}_2$ comprises $([x]_2, [e]_i, [s]_i, [y]_2)$. The simulator $S$ generates simulated views for $\mathcal{P}_1$ and $\mathcal{P}_2$, which are uniformly and randomly distributed. Computationally speaking, the adversary $A$ is unable to distinguish between the real and simulated views. Furthermore, leveraging $\operatorname{Lemma 1-3}$ ensures that the security guarantees provided by already proven protocols extend to the remaining protocols as sub-protocols. This demonstrates their overall security. Specifically in the case of the $\operatorname{SecNMS}$ protocol, parameters such as bounding box width, height, and area are exchanged between $\mathcal{P}_1$ and $\mathcal{P}_2$. However, this information alone does not allow adversary $A$ to infer the position of these bounding boxes; thus ensuring that coordinate values remain undisclosed.

\textbf{  Theorem 3:} Within the HBC model, the $\operatorname{SecYOLOv7}$ image processing and image decryption are secure.

$\operatorname{SecYOLOv7}$ encompasses several stages  incorporating the fundamental protocols mentioned in $\operatorname{Theorem 2}$, ensures the security of layers such as CBS, MP, E-ELAN, SPPCSPC, bounding-box prediction and classification. So the security is ensured.

\subsection{Complexity}

\begin{table}[ht]
    \centering
    \begin{tabular}{c|c|c}
        \hline
        Sec- & YOLOv7  & SecYOLOv7 \\
        \hline
        Conv  & $O(N)$   & $O(N)$        \\
        BN  & $O(N)$   & $O(N)$   \\
        Silu  & $O(N)$  & $O(N)$    \\
        MP  & $O(N/4)$  & $O(N/4)$ \\
        BBPred  & $O(N)$   & $O(N)$    \\
        DS  & $O(NlogN)$  & $O(NlogN)$      \\
        NMS  & $O(NlogN)$  & $O(NlogN)$      \\
        \hline
    \end{tabular}
    \caption{Comparison Of Computational Complexity}
\end{table}

\begin{table}[ht]
\begin{tabular}{c|c|c|c|c|c}
\hline
\multirow{2}{*}{Sec-} & \multicolumn{2}{c|}{SecYOLOv7} & \multirow{2}{*}{Sec-} & \multicolumn{2}{c}{SecYOLOv7} \\ \cline{2-3} \cline{5-6} 
 & \multicolumn{1}{c|}{Rounds} & OH &  & \multicolumn{1}{c|}{Rounds} & OH \\
 \hline
Exp & $4$ & $6$ & Silu & $10$ & $16N-8$ \\
Divi & $4$ & $\frac{7N-7}{2}$ & MP & $N^2-1$ & $4N^2-4$ \\
Comp & $8$ & $16$ & BBPred & $10$ & $6N+18$ \\
Conv & $\frac{N-1}{2}$ & $\frac{2N-1}{2}$ & DS & $1$ & $2N$ \\
BN & $0$ & $0$ & NMS & $5$ & $3N-1$\\
\hline
\end{tabular}
\caption{Communication Complexity}
\end{table}

Assuming the length of the input array is $N$, the Conv layer consists solely of $\operatorname{SSAdd}$ and $\operatorname{GRRMult}$ operations. In the case of the $\operatorname{SecDS}$ protocol utilized by $\operatorname{YOLOv7}$, comparison protocols are employed multiple times resulting in a computational complexity akin to quicksort ($O(NlogN)$) for $\operatorname{SDS}$ protocol. Consequently, the calculation complexity for $\operatorname{SecNMS}$ protocol, executing $\operatorname{SDS}$ protocol iteratively while performing $IOU$ calculations—is also $O(NlogN)$.

TABLE \uppercase\expandafter{\romannumeral1} provides a comparison of the computational complexities of the secure computation protocols discussed in this paper, with $N$ representing the length of the input array. The proposed secure computation protocols exhibit comparable computational complexity to $\operatorname{YOLOv7}$. However, due to the utilization of Shamir's secret sharing in this study, $\operatorname{SecYOLOv7}$ requires additional communication rounds and overhead. TABLE \uppercase\expandafter{\romannumeral2} calculates the communication rounds and overhead.

\begin{figure*} [t!]
	\centering
    \setlength{\intextsep}{0pt}
	\subfloat[\label{fig:a}]{
		\includegraphics[scale=0.22]{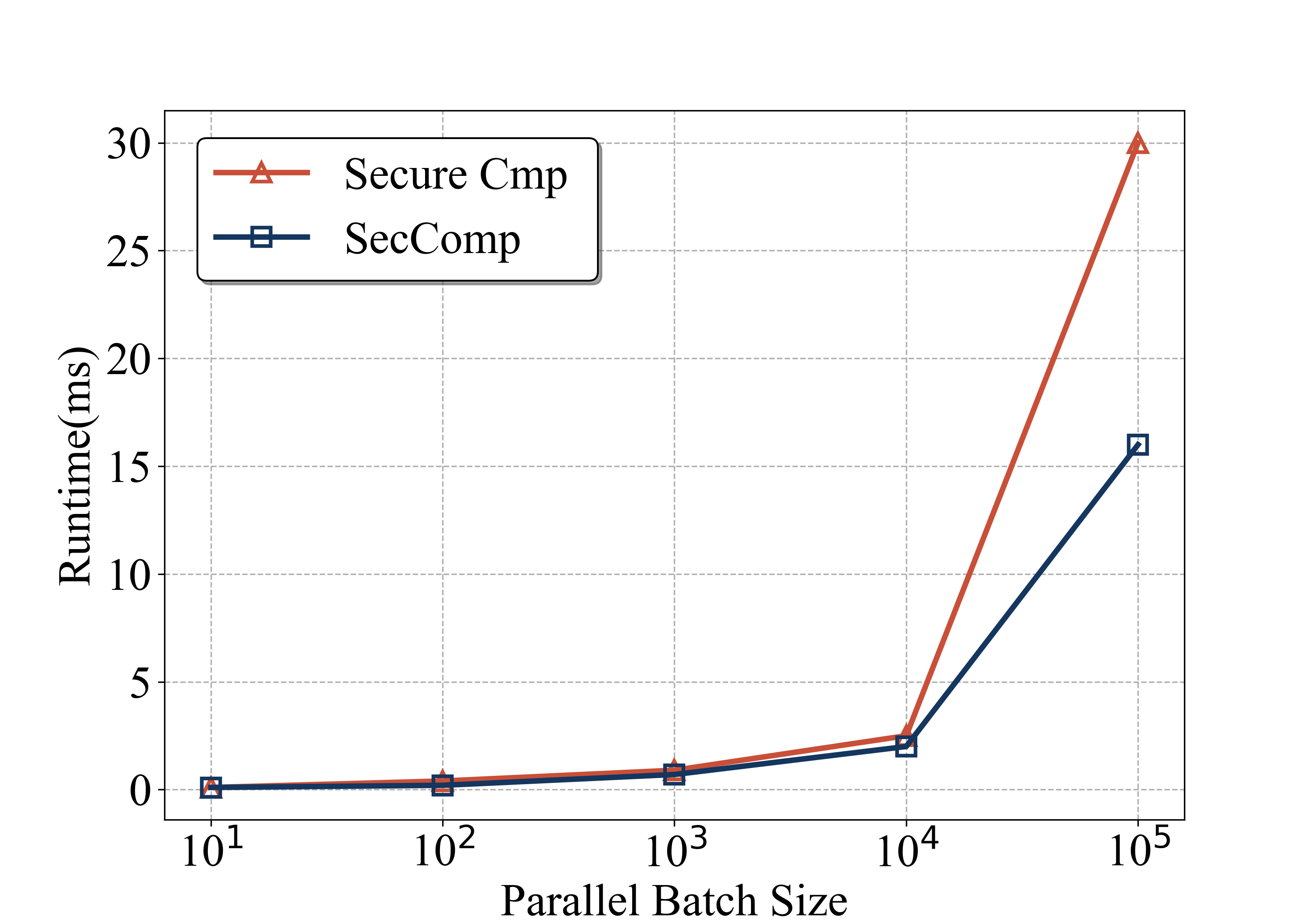}}
	\subfloat[\label{fig:b}]{
		\includegraphics[scale=0.22]{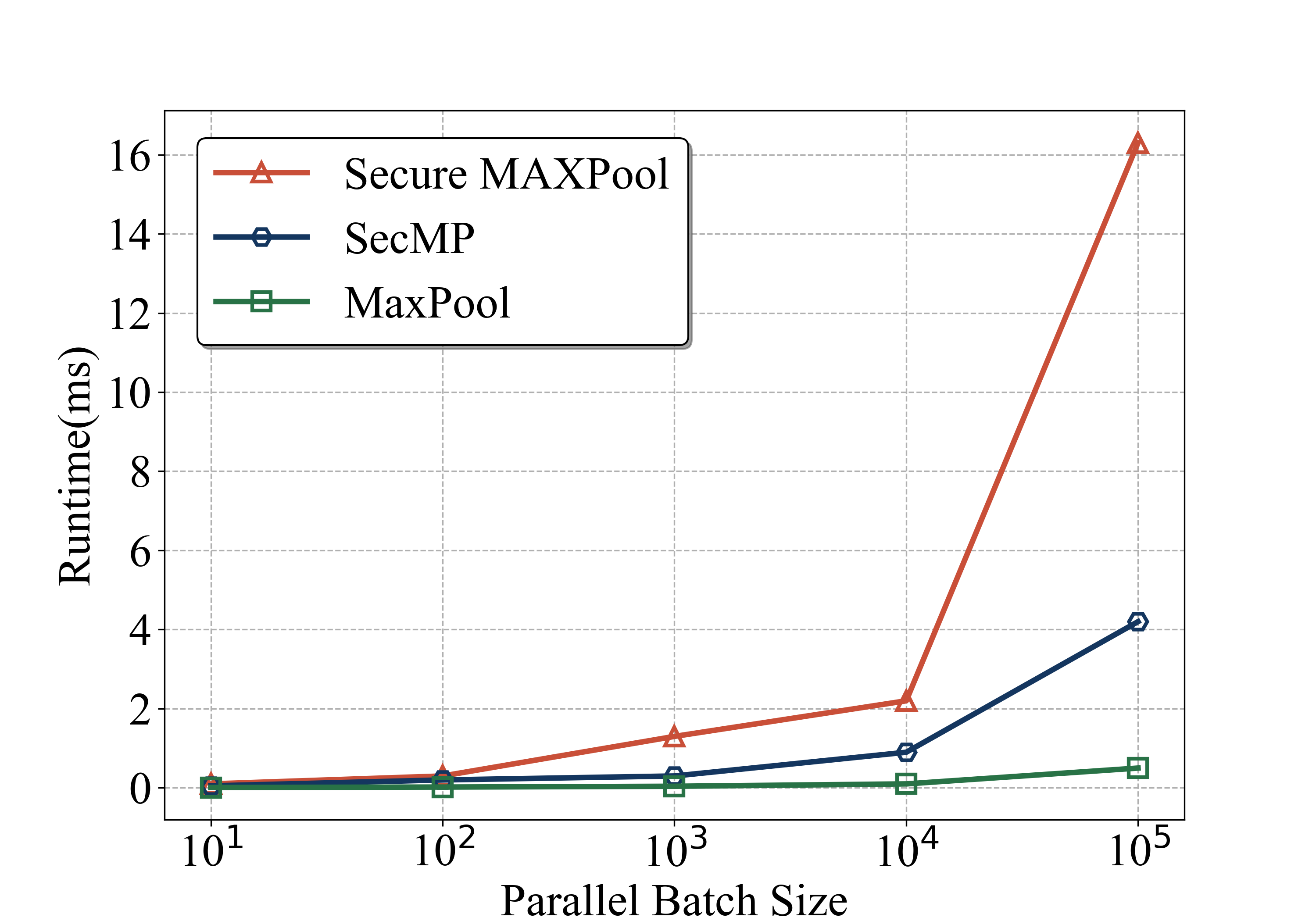}}
	\subfloat[\label{fig:c}]{
		\includegraphics[scale=0.22]{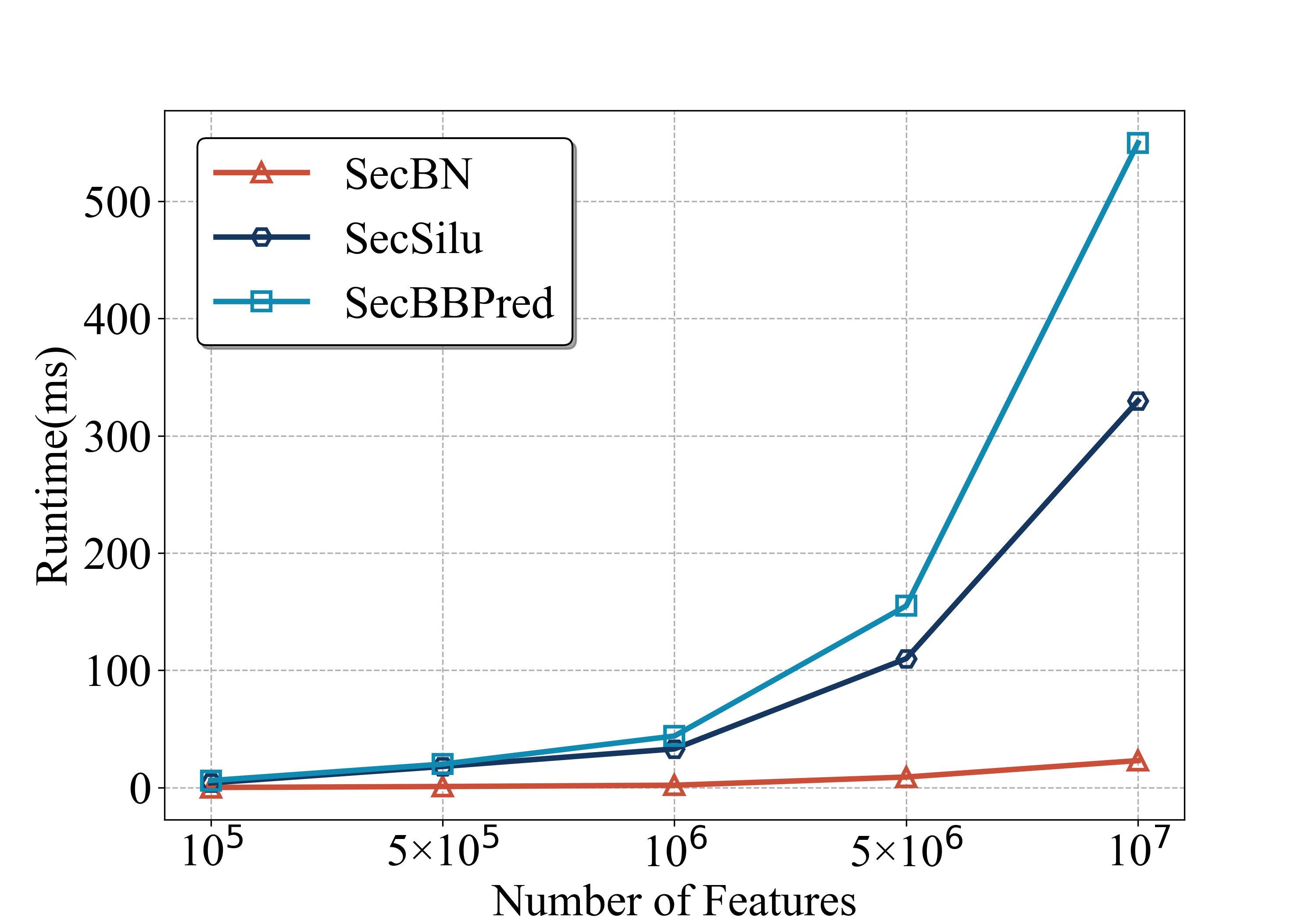}}
	\\
	\subfloat[\label{fig:d}]{
		\includegraphics[scale=0.22]{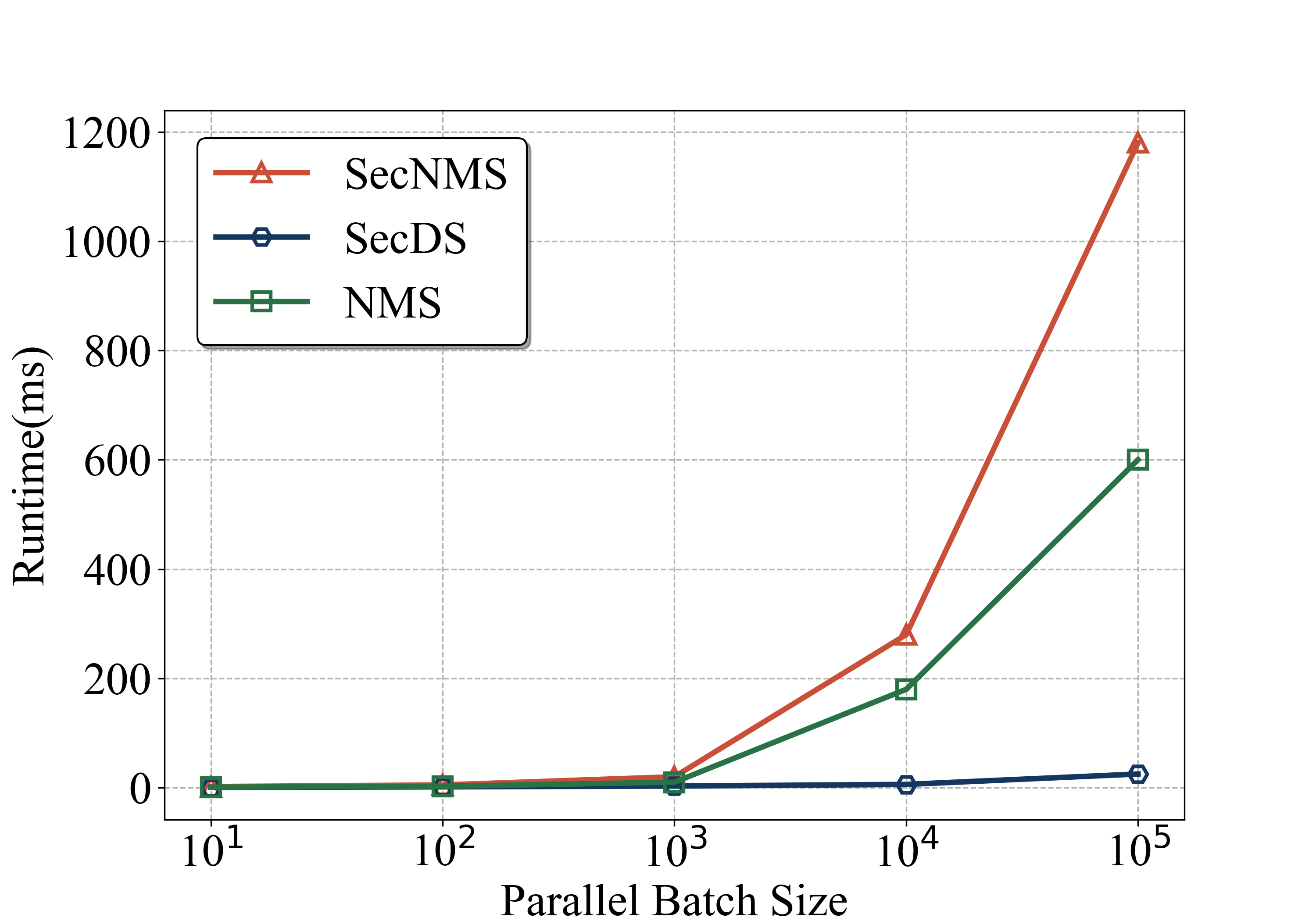} }
	\subfloat[\label{fig:e}]{
		\includegraphics[scale=0.22]{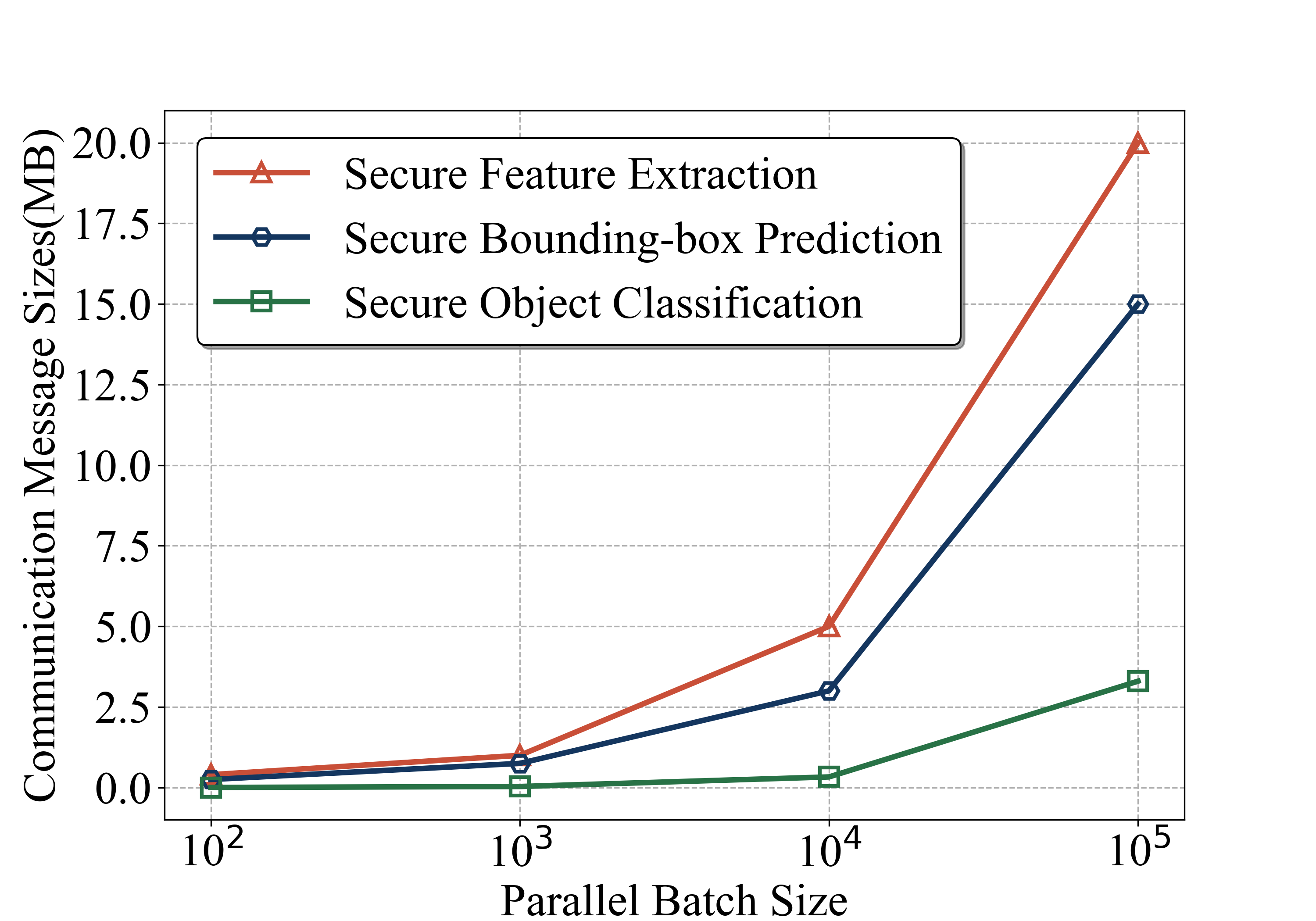}}
	\subfloat[\label{fig:f}]{
		\includegraphics[scale=0.22]{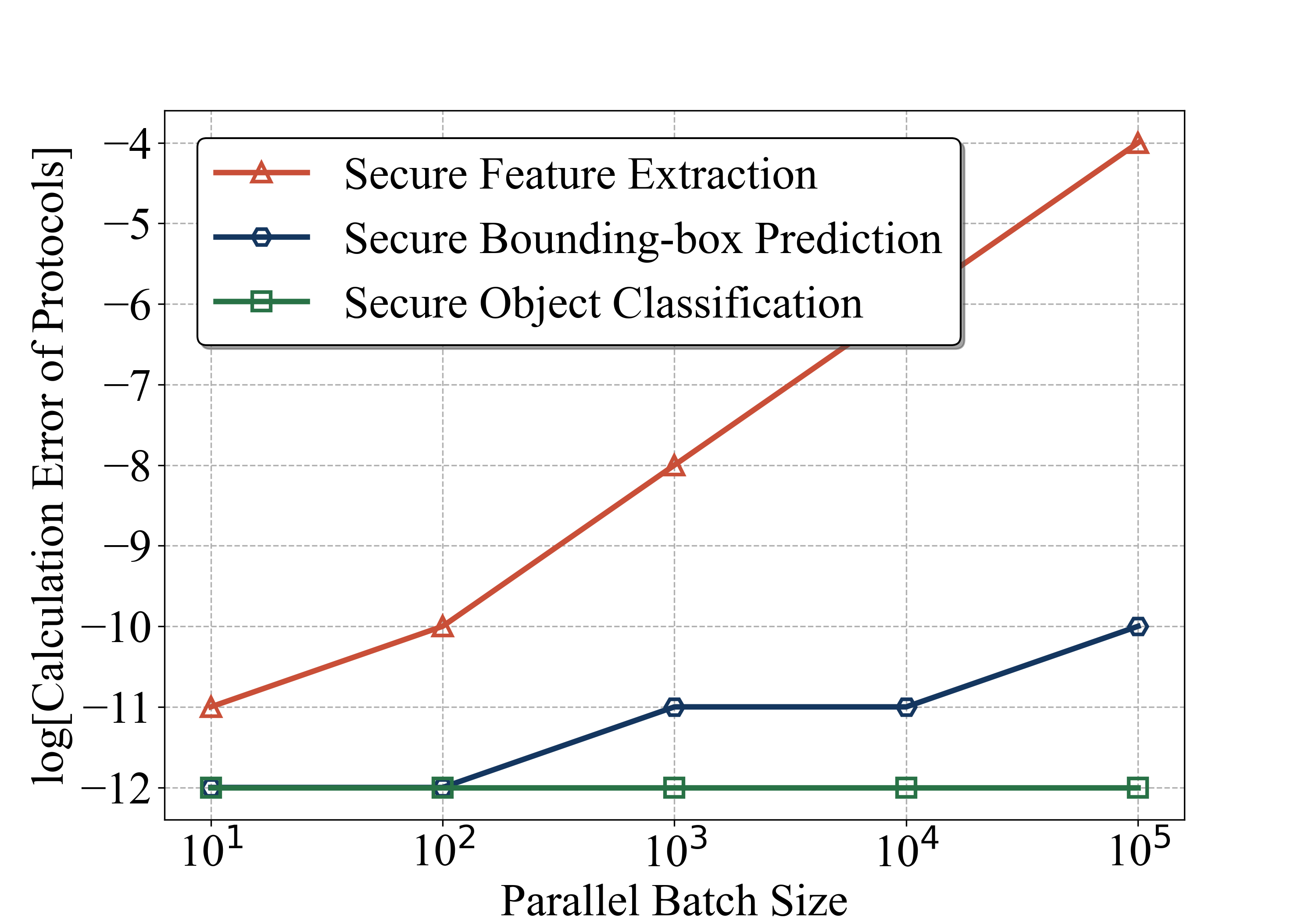}}
	\caption{Performance of proposed computing protocols (a) Comparison of computational costs between SecComp and Secure Cmp[21], (b) Comparison of computational costs between SecMP and Secure MAXPool[21], and MaxPool in plaintext environment, (c) Computational costs of SecBN, SecSilu, SecBBPred (d) Computational costs of SecNMS, SecDS, and NMS in plaintext environment (e) communication overheads of the three main modules (f) Calculation Errors of the three main modules}
	\label{fig3} 
\end{figure*}

\section{Experiments and Performance Evaluation}

\subsection{Experiment Settings}
We trained the $\operatorname{SecYOLOv7}$ network separately on plaintext and encrypted images. All experiments were conducted on a computer running Linux, with hardware configuration including 22 vCPU AMD EPYC 7T83 64-Core Processor CPU, Nvidia GeForce RTX 4090 GPU, 80GB of RAM, 24GB of memory, AutoDL as the platform, and Python 3.7 as the interpreter. Specifically, we used the pycryptodome library to simulate a trusted third party $\mathcal{T}$, for generating random keys.

In this experiment, we employed a fixed-point number representation similar to \cite{20}. To encode a fixed-point number $\widetilde{x}$ with f-bit precision, we map it to an integer $\bar{x}$. The original secure linear secret sharing protocols \cite{30} only support secure computations on integers. However, the training process of the model often involves floating-point numbers. To avoid precision loss (e.g., for the SecExp protocol), we truncate a fixed-point number $\bar{z}$ to $ \bar{z}^{\prime} $ using the Trunc protocol \cite{31}. All random shares are in the (1) format.

\newcommand{\highlight}[2]{\colorbox{#1!17}{$\displaystyle #2$}}
\newcommand{\highlightdark}[2]{\colorbox{#1!47}{$\displaystyle #2$}}
\begin{equation}
\vspace{\baselineskip}
\label{eq:epsilon}
\tikzmarknode{z}{\highlight{green}{\bar{z}^{\prime}}}=\tikzmarknode{x}{\highlight{yellow}{\widetilde{x}* \widetilde{y} * 2^f}}
\end{equation}
\begin{tikzpicture}[overlay,remember picture,>=stealth,nodes={align=left,inner ysep=1pt},<-]
\path (z.south) ++ (0,-1em) node[anchor=north east,color=red!67] (scalep){$ \text{for }  \bar{z}=\widetilde{x}* \widetilde{y} * {2^{2f}} $};
\draw [color=red!87](z.south) |- ([xshift=-0.3ex,color=red]scalep.south west);

\path (x.south) ++ (0,-1em) node[anchor=north west,color=blue!67] (mean){\text{fixed-point }$\bar{x}=\widetilde{x} * 2^f$};
\draw [color=blue!57](x.south) |- ([xshift=-0.3ex,color=blue]mean.south east);
\end{tikzpicture}
\subsection{Performance of Protocols}

We assess the computational efficiency of the key computational protocols mentioned in this paper by measuring the computation time of a specific protocol through breakpoint testing. Figures 3(a) and 3(b) demonstrate that the runtime of both $\operatorname{SecComp}$ and $\operatorname{SecMP}$ protocols increases with an increase in Parallel Batch Size. Our proposed $\operatorname{SecComp}$ protocol only uses binary linear operation circuits, resulting in approximately $1/3$ cost savings compared to the $\operatorname{SecureCmp}$ protocol proposed in \cite{19}, which utilizes beaver triple operator. Figure 3(c) shows that dividing feature maps into sub-regions at Max-Pool layer results in better computational cost performance for our proposed $\operatorname{SecMP}$ protocol than \cite{19}. The runtime of our proposed $\operatorname{SecBN}$, $\operatorname{SecSilu}$, and $\operatorname { SecBBPred }$ protocols increases with an increase in number of features but performs well when dealing with large arrays. However, Figure 3(d) demonstrates that processing an array of length $10^5$ using our proposed $\operatorname { SecNMS }$ protocol incurs relatively high time cost taking around 1200ms.

We assessed communication overhead and computational errors of three main modules of $\operatorname{SecYOLOv7}$ in Figure 3(e). The amount of data transmitted between servers can be used to measure the communication overhead of the workload. The number and range of inputs respectively affect runtime and error rate. When processing arrays with a length of $10^5$, both Secure Feature Contraction and Secure Bounding-Box Prediction result in relatively large communication overheads, but they are both kept within 20MB. Our proposed protocols outperform those presented in \cite{19} regarding their communication overhead.

We further quantify the computational error as the maximum deviation between the output results of secure computation protocols and plaintext functions. Interestingly, due to the $\operatorname{SecComp}$ protocol's insensitivity towards computation errors in the integer part and its disregard for sign bit influence on comparison results, we designed the $\operatorname{SecNMS}$ protocol based on this assumption. The Secure Object Classification embedded with the $\operatorname{SecNMS}$ protocol achieves nearly zero error, disregarding system jitter factors. Figure 3(f) illustrates that when processing arrays of length $10^5$, all protocols exhibit computational errors within a magnitude of $10^{-4}$.

\subsection{Performance of Transmission Line Detection}

\begin{figure*} [t!]
	\centering
	\subfloat[\label{fig:a}]{
		\includegraphics[scale=0.34]{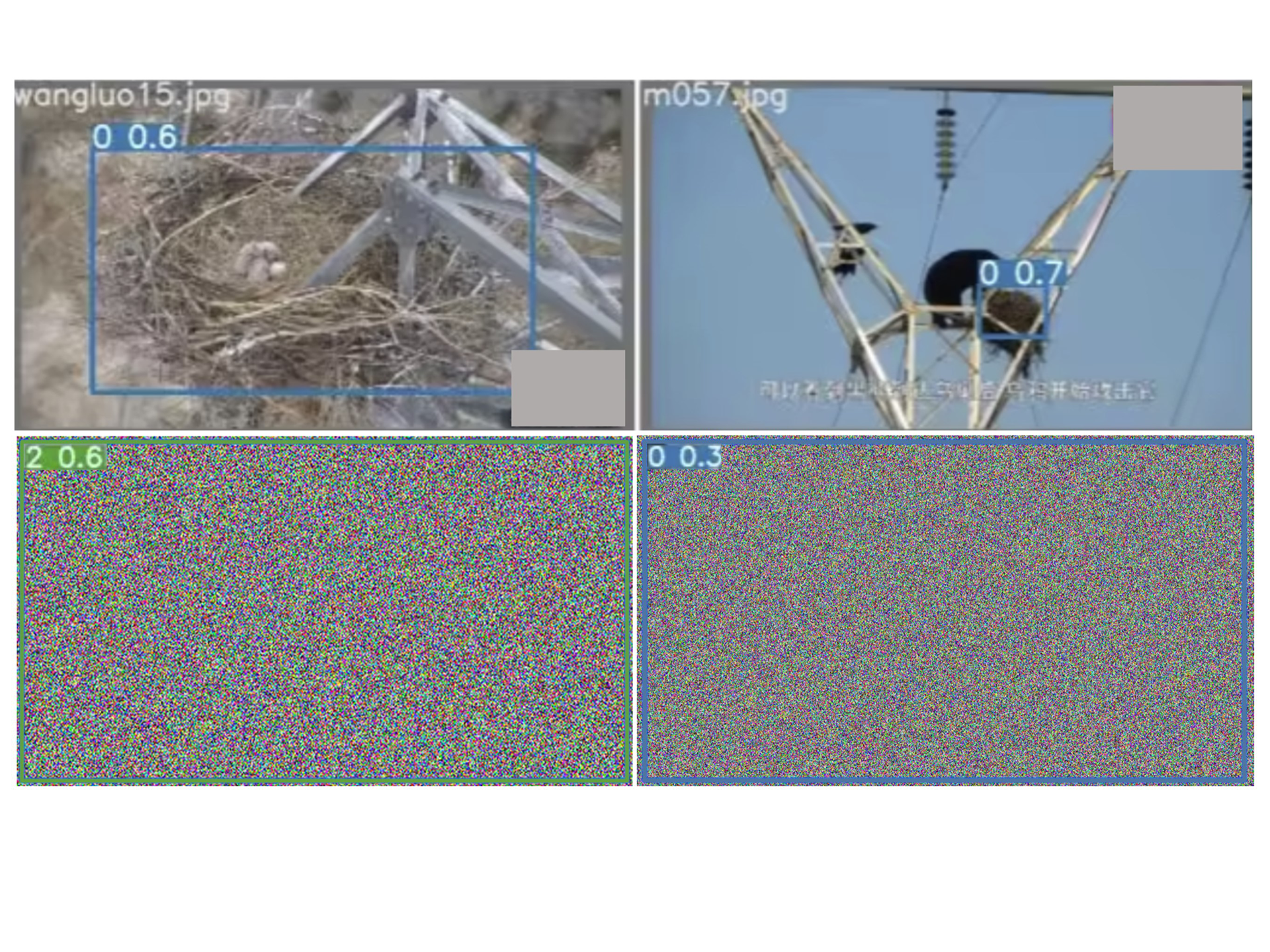}}
	\subfloat[\label{fig:b}]{
		\includegraphics[scale=0.3]{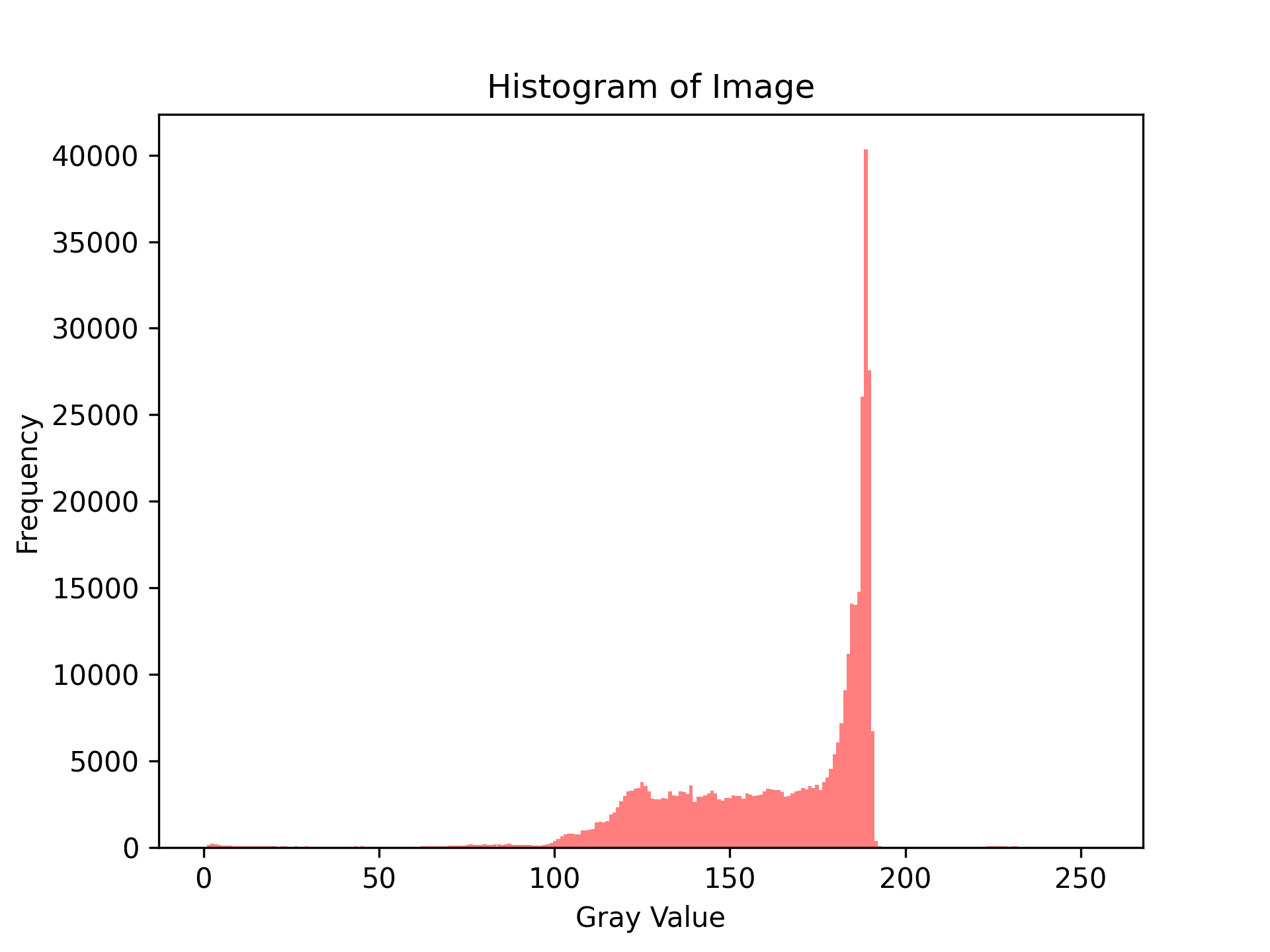}}
	\subfloat[\label{fig:c}]{
		\includegraphics[scale=0.3]{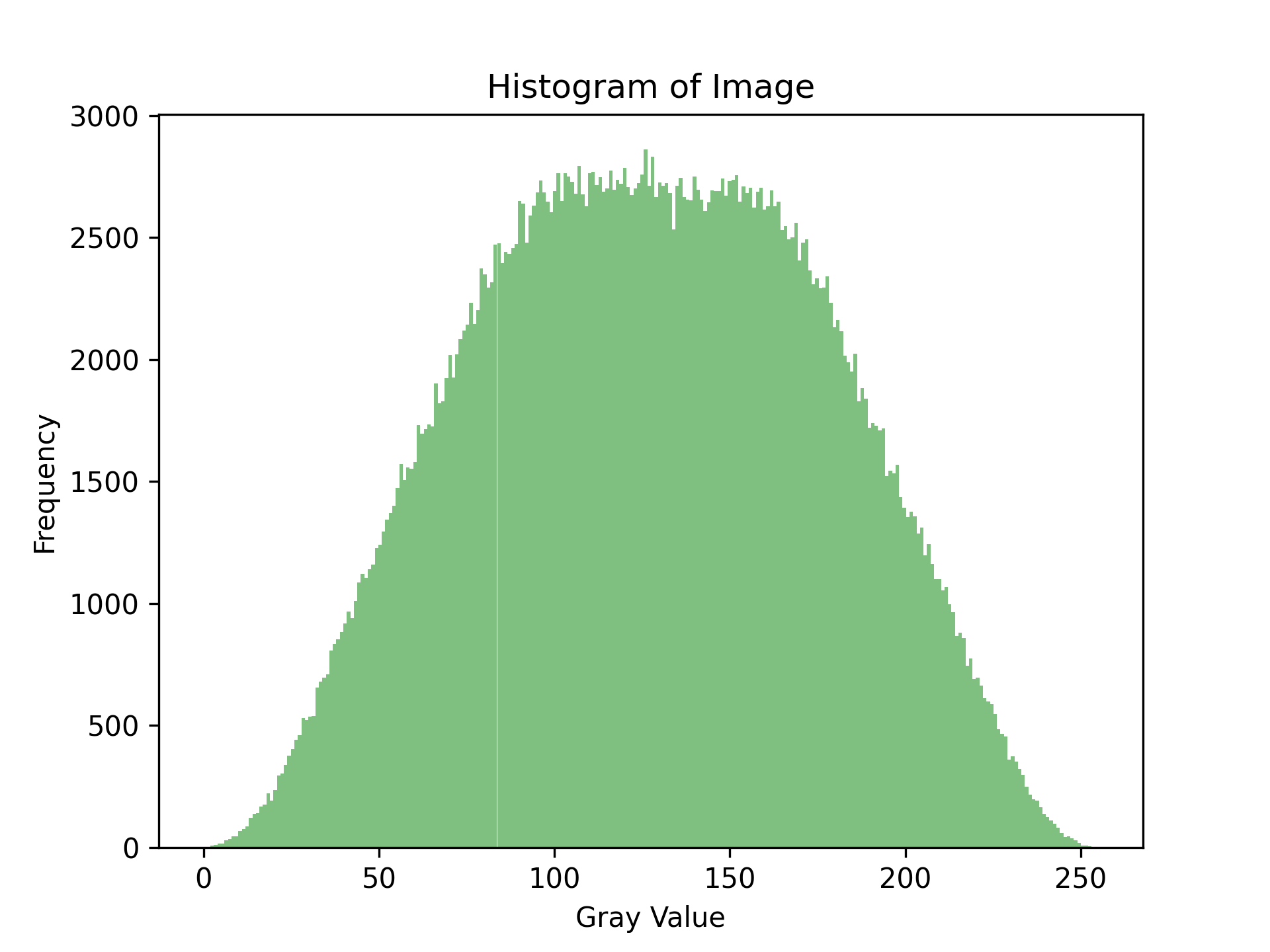}}
    \subfloat[\label{fig:d}]{
    		\includegraphics[scale=0.28]{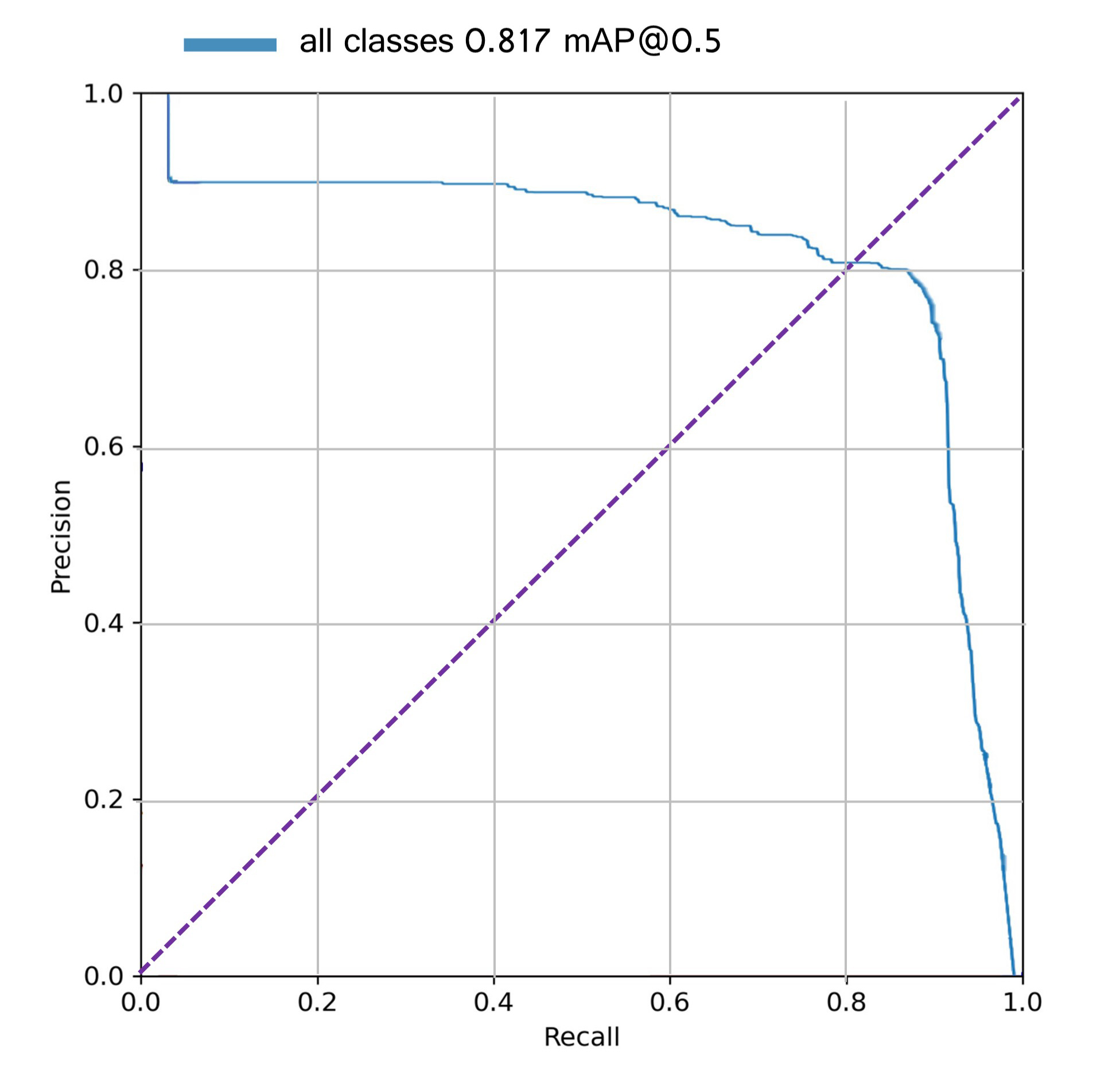}}
	\caption{Performance of Transmission line object detection (a) Prediction result of two original images and their secure shares in $\mathcal{P}_i$ (b) (c) Gray value histogram of original and secure images (d) P-R curve of SecYOLOv7}
	\label{fig4} 
\end{figure*}

We utilized the common foreign matter dataset of transmission line for experimentation, which includes four categories (Nest, Balloon, Kite and Trash) and a total of 4516 images (2297 training images and 2219 val images). The computational cost of $\operatorname{SecYOLOv7}$ was measured to be 2.113 s, approximately one-tenth that of $\operatorname{YOLOv7}$ in plaintext environment. Additionally, the communication overhead was found to be 95.15 MB, demonstrating improved performance compared to existing work. To execute $\operatorname{SecYOLOv7}$ on a randomly selected image $\mathcal{P}_i$, we employed a secure computing protocol designed for interactive execution using pixel-level image shares $I_i$ derived from original image $I$. In this experiment, $l = 8$ was set to ensure security given that each pixel value in $I_i$ falls within $[0-255]$. Fig4(b)(c) depict the pixel histograms of original picture $I$ and secret shared picture ($I_i$). Notably, gray values in $I$ are clustered within a certain range while those in $I_i$ are normally distributed between $[0-255]$; thus indicating that statistical distribution characteristics of original image have been concealed through our approach. Linear reconstruction can be used to recover the original image.

The detection result $O_i$ obtained after the collaboration of two pis to implement the $\operatorname{SecYOLOv7}$ framework is illustrated in Fig4(a). $O_i$ represents a share of object detection results, which lacks information about the class, shape, and position of the objects. From $\mathcal{P}_i$'s perspective, each detected object's boundary is expanded to encompass the entire image, leading to incorrect classification. It can be observed that both $\mathcal{P}_i$ and adversary $A$ fail to obtain accurate detection results under the given constraints, demonstrating the security of $\operatorname{SecYOLOv7}$. Furthermore, we train our model in a plaintext environment and perform inference on two random image shares. This experiment is repeated 300 times to update model parameters by minimizing classification and regression losses. As depicted in Fig4(d), based on recall and precision indicators along with the definition of balance point, an average precision value of $81.73\%$ can be calculated for all classes.

\section{Conclusion}

This paper presents secure computation protocols based on the Shamir's secret sharing scheme for Edge Collaboration. The protocols are organized into three modules: Secure Feature Contraction, Secure Bounding-Box Prediction, and Secure Object Classification. $\operatorname{SecYOLOv7}$ is constructed using $\operatorname{YOLOv7}$ to enable privacy-preserving object detection in the electric transmission line while protecting target features and locations. A theoretical analysis demonstrates the security and complexity of these protocols and framework. Experimental validation with an electric transmission line dataset shows that $\operatorname{SecYOLOv7}$ incurs low time cost, low communication overhead among edges, and a low calculation error. These findings have significant implications for real-time demanding applications such as UAV inspections.


\bibliography{1}

\begin{thebibliography}{10}
\providecommand{\url}[1]{#1}
\csname url@samestyle\endcsname
\providecommand{\newblock}{\relax}
\providecommand{\bibinfo}[2]{#2}
\providecommand{\BIBentrySTDinterwordspacing}{\spaceskip=0pt\relax}
\providecommand{\BIBentryALTinterwordstretchfactor}{4}
\providecommand{\BIBentryALTinterwordspacing}{\spaceskip=\fontdimen2\font plus
\BIBentryALTinterwordstretchfactor\fontdimen3\font minus
  \fontdimen4\font\relax}
\providecommand{\BIBforeignlanguage}[2]{{%
\expandafter\ifx\csname l@#1\endcsname\relax
\typeout{** WARNING: IEEEtran.bst: No hyphenation pattern has been}%
\typeout{** loaded for the language `#1'. Using the pattern for}%
\typeout{** the default language instead.}%
\else
\language=\csname l@#1\endcsname
\fi
#2}}
\providecommand{\BIBdecl}{\relax}
\BIBdecl

\bibitem{1}
Y.~L. Z. D. G. W. L.~L. Zhang, ``Uav-aided urban target tracking system based
  on edge computing,'' in \emph{CoRR}, 2019.

\bibitem{3}
C.~Zhang, X.~Liu, B.~Chen, P.~Yin, J.~Li, Y.~Li, and X.~Meng, ``Insulator
  profile detection of transmission line based on traditional edge detection
  algorithm,'' in \emph{2020 IEEE International Conference on Artificial
  Intelligence and Computer Applications (ICAICA)}, 2020.

\bibitem{2}
X.~Cao, J.~Xu, and R.~Zhang, ``Mobile edge computing for cellular-connected
  uav: Computation offloading and trajectory optimization,'' in \emph{2018 IEEE
  19th International Workshop on Signal Processing Advances in Wireless
  Communications (SPAWC)}, 2018.

\bibitem{4}
M.~Lan, Y.~Zhang, L.~Zhang, and B.~Du, ``Defect detection from uav images based
  on region-based cnns,'' in \emph{2018 IEEE International Conference on Data
  Mining Workshops (ICDMW)}, 2018.

\bibitem{5}
F.~Li, J.~Xin, T.~Chen, L.~Xin, Z.~Wei, Y.~Li, Y.~Zhang, H.~Jin, Y.~Tu,
  X.~Zhou, and H.~Liao, ``An automatic detection method of bird’s nest on
  transmission line tower based on faster rcnn,'' \emph{IEEE Access}, 2020.

\bibitem{6}
Z.~ZHAO, Y.~LI, Z.~ZHEN, Y.~ZHAI, K.~ZHANG, and W.~ZHAO, ``Typical fittings
  detection method with faster r-cnn combining kl divergence and shape
  constraints,'' \emph{High Voltage Engineering}, 2020.

\bibitem{7}
X.~Tao, D.~Zhang, Z.~Wang, X.~Liu, H.~Zhang, and D.~Xu, ``Detection of power
  line insulator defects using aerial images analyzed with convolutional neural
  networks,'' \emph{IEEE Transactions on Systems, Man, and Cybernetics:
  Systems}, 2020.

\bibitem{8}
L.~Cui, Y.~Qu, G.~Xie, D.~Zeng, R.~Li, S.~Shen, and S.~Yu, ``Security and
  privacy-enhanced federated learning for anomaly detection in iot
  infrastructures,'' \emph{IEEE Transactions on Industrial Informatics}, 2022.

\bibitem{9}
Z.~Wei-jing, Z.~He-chun, Y.~Shi-ying, and L.~Tong, ``A homomorphic
  encryption-based privacy preserving data aggregation scheme for smart grid,''
  in \emph{15th CIS}, 2019.

\bibitem{10}
R.~Qiu, M.~Ai, F.~Zheng, L.~Liang, and Y.~Li, ``Privacy-preserving of power
  consumption big data based on improved group signature and homomorphic
  encryption,'' in \emph{IEEE 3rd AUTEEE}, 2020.

\bibitem{11}
S.~Ren, K.~He, R.~Girshick, and J.~Sun, ``Faster r-cnn: Towards real-time
  object detection with region proposal networks,'' \emph{IEEE Transactions on
  Pattern Analysis and Machine Intelligence}, 2017.

\bibitem{12}
Y.~Tang, J.~Han, W.~Wei, J.~Ding, and X.~Peng, ``Research on part recognition
  and defect detection of transmission line in deep learning,'' 2017.

\bibitem{122}
Z.~Hu, ``Intelligent identification of hidden troubles in key parts of
  transmission towers based on edge computing and uav inspection images,''
  2020.

\bibitem{13}
J.~Redmon, S.~Divvala, R.~Girshick, and A.~Farhadi, ``You only look once:
  Unified, real-time object detection,'' in \emph{CVPR}, 2016.

\bibitem{14}
W.~Liu, D.~Anguelov, D.~Erhan, C.~Szegedy, S.~Reed, C.-Y. Fu, and A.~C. Berg,
  ``Ssd: Single shot multibox detector,'' in \emph{Computer Vision -- ECCV
  2016}, B.~Leibe, J.~Matas, N.~Sebe, and M.~Welling, Eds., 2016.

\bibitem{15}
H.~Ohta, Y.~Sato, T.~Mori, K.~Takaya, and V.~Kroumov, ``Image acquisition of
  power line transmission towers using uav and deep learning technique for
  insulators localization and recognition,'' in \emph{23rd ICSTCC}, 2019.

\bibitem{155}
Y.~Huang, Z.~Chen, Q.~Chen, L.~Zhang, H.~Liu, and J.~Zhang, ``Real-time
  detection method for transmission line faults applying edge computing and
  improved yolov5s,'' 2023.

\bibitem{16}
S.~Moriai, ``Privacy-preserving deep learning via additively homomorphic
  encryption,'' in \emph{2019 IEEE 26th Symposium on Computer Arithmetic
  (ARITH)}, 2019.

\bibitem{17}
C.~Juvekar, V.~Vaikuntanathan, and A.~Chandrakasan, ``Gazelle: A low latency
  framework for secure neural network inference,'' 2018.

\bibitem{18}
F.~McSherry and K.~Talwar, ``Mechanism design via differential privacy,'' in
  \emph{48th Annual IEEE Symposium on Foundations of Computer Science
  (FOCS'07)}, 2007.

\bibitem{19}
K.~Huang, X.~Liu, S.~Fu, D.~Guo, and M.~Xu, ``A lightweight privacy-preserving
  cnn feature extraction framework for mobile sensing,'' \emph{IEEE
  Transactions on Dependable and Secure Computing}, 2021.

\bibitem{20}
W.~Ruan, M.~Xu, W.~Fang, L.~Wang, L.~Wang, and W.~Han, ``Private, efficient,
  and accurate: Protecting models trained by multi-party learning with
  differential privacy,'' in \emph{IEEE SP)}, 2023.

\bibitem{21}
C.-Y. Wang, A.~Bochkovskiy, and H.-Y.~M. Liao, ``{YOLOv7}: Trainable
  bag-of-freebies sets new state-of-the-art for real-time object detectors,''
  \emph{arXiv preprint arXiv:2207.02696}, 2022.

\bibitem{23}
A.~Shamir, ``How to share a secret,'' 1979.

\bibitem{24}
R.~Cramer, I.~Damg{\aa}rd, and Y.~Ishai, ``Share conversion, pseudorandom
  secret-sharing and applications to secure computation,'' in \emph{Theory of
  Cryptography}, 2005.

\bibitem{25}
R.~Gennaro, M.~O. Rabin, and T.~Rabin, ``Simplified vss and fast-track
  multiparty computations with applications to threshold cryptography,'' in
  \emph{Proceedings of the Seventeenth Annual ACM Symposium on Principles of
  Distributed Computing}, 1998.

\bibitem{26}
D.~Bogdanov, S.~Laur, and J.~Willemson, ``Sharemind: A framework for fast
  privacy-preserving computations,'' in \emph{13th European Symposium on
  Research in Computer Security}, 2008.

\bibitem{27}
D.~Bogdanov, M.~Niitsoo, T.~Toft, and J.~Willemson, ``High-performance secure
  multi-party computation for data mining applications,'' \emph{Int. J. Inf.
  Secur.}, 2012.

\bibitem{28}
D.~Feng and K.~Yang, ``Concretely efficient secure multi-party computation
  protocols: survey and more,'' \emph{Security and Safety}, 2022.

\bibitem{30}
M.~Ben-Or, S.~Goldwasser, and A.~Wigderson, ``Completeness theorems for
  non-cryptographic fault-tolerant distributed computation,'' in
  \emph{Proceedings of the Twentieth Annual ACM Symposium on Theory of
  Computing}, 1988.

\bibitem{31}
O.~Catrina, ``Round-efficient protocols for secure multiparty fixed-point
  arithmetic,'' in \emph{2018 International Conference on Communications
  (COMM)}, 2018.

\end{thebibliography}
\bibliographystyle{IEEEtran}



%

\end{document}